\documentclass[prd,twocolumn,groupedaddress,nofootinbib,english,floatfix,superscriptaddress,preprintnumbers]{revtex4-1}
\usepackage{graphicx}
\usepackage{cancel}
\usepackage{amssymb}
\usepackage{textcomp}
\usepackage{amsmath}
\usepackage{bm}
\usepackage{times}
\usepackage{epsfig}
\usepackage{graphics}
\usepackage{setspace}
\usepackage{comment}
\usepackage{subcaption}
\usepackage{color}
\definecolor{naviBlue}{RGB}{0,0,128}
\usepackage[colorlinks  = true,
            linkcolor   = naviBlue,
            urlcolor    = naviBlue,
            citecolor   = naviBlue,
            anchorcolor = naviBlue]{hyperref}

\usepackage[english]{babel}
\usepackage{textcomp}
\usepackage{braket}
\usepackage{siunitx}
\usepackage{latexsym}
\usepackage{mathrsfs}
\usepackage[a4paper, left=2.5cm, right=2.5cm, top=2.5cm]{geometry}

\usepackage{tikz}
\usetikzlibrary{patterns}
\usetikzlibrary{shapes.geometric}
\usetikzlibrary{decorations.pathmorphing}
\usetikzlibrary{arrows.meta}
\tikzset{snake it/.style={decorate, decoration=snake}}
\usetikzlibrary{positioning}
\usetikzlibrary{arrows,shapes,positioning}
\usetikzlibrary{decorations.markings}
\tikzstyle arrowstyle=[scale=1]
\tikzstyle directed=[postaction={decorate,decoration={markings,mark=at position .65 with {\arrow[arrowstyle]{stealth}}}}]
\tikzstyle reverse directed=[postaction={decorate,decoration={markings,mark=at position .65 with {\arrowreversed[arrowstyle]{stealth};}}}]

\newcommand{\rC}{{\mathrm{C}}}
\newcommand{\rU}{{\mathrm{U}}}
\newcommand{\rI}{{\mathrm{I}}}
\newcommand{\rII}{{\mathrm{II}}}
\newcommand{\rIII}{{\mathrm{III}}}

\begin{document}
\title{Quantum stress tensor at the Cauchy horizon of Reissner-Nordstr\"om-deSitter spacetime}

\author{Stefan Hollands}
\email{stefan.hollands@uni-leipzig.de}
\author{Christiane Klein}
\email{klein@itp.uni-leipzig.de}
\author{Jochen Zahn}
\email{jochen.zahn@itp.uni-leipzig.de}
\affiliation{Institut f\"ur Theoretische Physik, Universit\"at Leipzig,\\ Br\"uderstra{\ss}e 16, 04103 Leipzig, Germany}

\begin{abstract}
The strong cosmic censorship conjecture proposes that starting from generic initial data on some Cauchy surface, the solutions of the Einstein equation should not be extendable across the boundary of the domain of dependence of  that surface. For the case of the Reissner-Nordstr\"om-de Sitter spacetime this means that any perturbation should blow up sufficiently badly when approaching this boundary, called the Cauchy horizon. However, recent results indicate that for highly charged black holes classical scalar perturbations allow for a violation of strong cosmic censorship. In a recent paper \cite{Hollands:2019}, two of us have argued that quantum effects will restore censorship for generic values of the black hole parameters. But, due to practical limitations, the precise form of the divergence was only calculated for a small number of parameters. Here we perform a thorough parameter scan using an alternative, more efficient semi-analytic method.
Our analysis confirms \cite{Hollands:2019} in the sense that the quantum stress tensor is found to diverge badly generically. However, the sign of the divergence can be changed by changing the mass of the field or the spacetime parameters, leading to a drastically different type of singularity on the Cauchy horizon.
\end{abstract}
\maketitle

\section{Introduction}
\begin{figure}
   \centering
\includegraphics[width=0.4\textwidth]{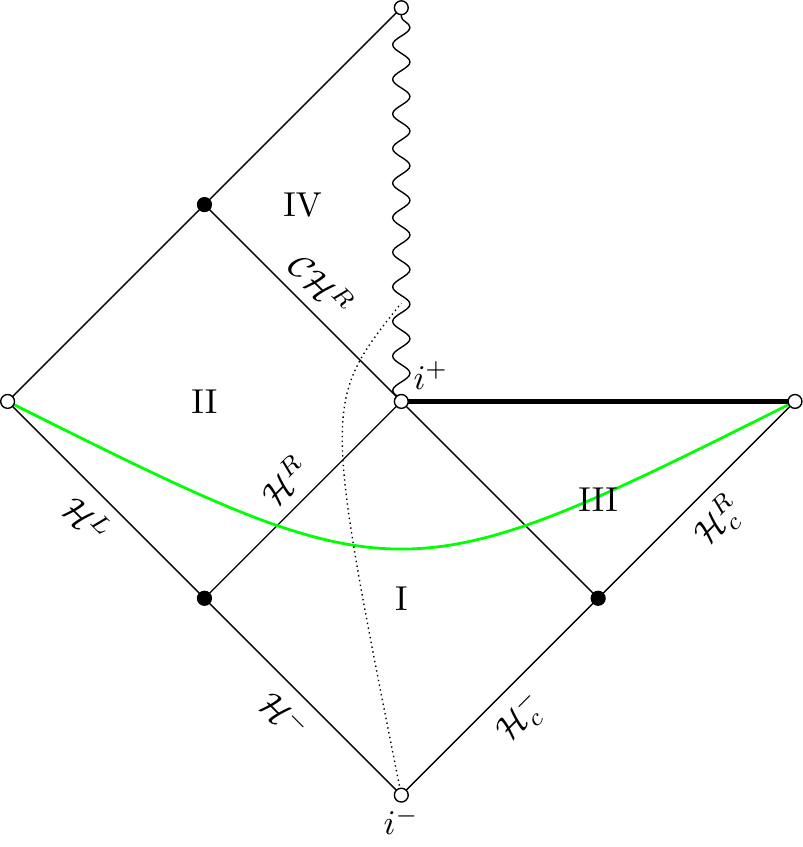}
\caption{\footnotesize Penrose diagram for the  Reissner-Nordstr\"om-de Sitter (RNdS) spacetime. The wiggled line represents the curvature singularity, the double lines correspond to conformal infinity. The other lines represent different horizons. Filled dots stand for bifurcation surfaces, while empty dots indicate singular points or points at infinity. The green line indicates a possible Cauchy surface for the region $\rI \cup \rII \cup \rIII$. The dotted line represents the surface of the collapsing matter in a typical collapse situation: the region on the right hand side of the line is outside of the collapsing matter.}
\label{fig:RNdS}
\end{figure}

The Reissner-Nordstr\"om-(de Sitter) RN(dS) family of spacetimes describes a static, spherically symmetric charged black hole with vanishing (positive) cosmological constant. The Penrose  diagram for the de Sitter-case is shown in figure~\ref{fig:RNdS}.

It is clear from this diagram that solutions for a generic hyperbolic equation with prescribed initial data on the Cauchy surface in green are determined uniquely only up to the 
``Cauchy horizon'', $\mathcal{CH}^R$. One may thus wonder whether a given solution could be continued in several ways (if at all) beyond this 
horizon. If so, this would physically represent a breakdown of determinism in General Relativity. The strong cosmic censorship (sCC) conjecture, here in a formulation due to Christodoulou \cite{Christodoulou:2008}, proposes that, for generic smooth initial data, the metric should in fact {\em not} be extendable across $\mathcal{CH}^R$ as a weak solution to the Einstein equations in the local Sobolev space $H^1_{\mathrm{loc}}$. It should also apply to perturbations by a scalar field satisfying the Klein-Gordon equation
\begin{align}
\label{eq:kge}
\left(\Box_g-\mu^2\right)\Phi=0\, ,
\end{align}
which can be considered as a toy model for metric perturbations \cite{Dias:2018}. The sCC conjecture is based on a heuristic argument by Penrose \cite{Penrose:1974}, who observed that perturbations approaching $\mathcal{CH}^R$ will be infinitely blue-shifted, leading to a divergent stress-energy tensor.

There have been numerous studies on sCC on RN(dS) spacetimes, e.g.\cite{Poisson:1989, Poisson:1990, Brady:1998, Mellor:1989, Mellor:1992, Dafermos:2003a,Dafermos:2003b, Dafermos:2012, Costa:2014a, Costa:2014b, Costa:2014c, Franzen:2014, Luk:2015, Hintz:2015, Costa:2016, Cardoso:2017, Hod:2018, Dias:2018, Dias:2018a, Cardoso:2018, Mo:2018,  Gim:2019, Zilberman:2019}. While in the case of vanishing cosmological constant e.g. the results by \cite{Luk:2015} indicate that sCC holds in the form stated above, the situation is more difficult and interesting when the constant is nonzero. In the presence of a positive cosmological constant, the redshift effect from cosmological expansion counteracts the blueshift effect, thus potentially providing a physical mechanism to violate sCC. In fact, the analytic results by \cite{Hintz:2015} combined with the numerical studies by \cite{Cardoso:2017, Dias:2018, Cardoso:2018} indicate that sCC {\em is} violated for strongly charged de Sitter black holes when taking into account classical perturbations.

Two of us therefore asked in \cite{Hollands:2019} whether sCC could be rescued by quantum effects in RNdS spacetimes.  For this, the expected energy flux towards the Cauchy horizon, $\langle T_{V V}\rangle_\Psi$, of a conformally coupled scalar quantum field was studied, 
where $V$ is a regular Kruskal-type coordinate vanishing on the Cauchy horizon $\mathcal{C}\mathcal{H}^R$.
It was shown that for {\em any} (!) state $\Psi$ which is Hadamard in the neighborhood of some Cauchy surface $\Sigma$ of $\rI \cup \rII \cup \rIII$
such as the green surface indicated in figure~\ref{fig:RNdS}, the blow up of  $\langle T_{V V}\rangle_\Psi$ near the Cauchy horizon is dominated by $\langle T_{V V}\rangle_\rU-\langle T_{V V}\rangle_\rC$, where the subscript $\rU$ denotes the Unruh vacuum and $\rC$ is a comparison state, which has been constructed explicitly in \cite{Hollands:2019}, and was proven to be Hadamard in the union of region $\rII$ and a two-sided neighbourhood of the Cauchy horizon. More precisely, it was shown that $\langle T_{V V}\rangle_\Psi$ splits into a ``classical'' piece not more divergent than the stress-energy tensor of the classical field, and the piece 
$\langle T_{V V}\rangle_\rU-\langle T_{V V}\rangle_\rC \sim \kappa_-^{-2} \tilde{C}V^{-2}$
where $ \kappa_-$ is the surface gravity on the Cauchy horizon, and $\tilde{C}$ is a constant which needs to be determined semi-analytically.
The $V^{-2}$ divergence of this piece is always worse than that of the
``classical'' piece provided of course that $\kappa_-^{-2} \tilde{C} \neq 0$, and in this case
\begin{equation}
\label{eq:TVVdiv}
\langle T_{V V}\rangle_\Psi \sim
\langle T_{V V}\rangle_\rU-\langle T_{V V}\rangle_\rC \sim \kappa_-^{-2} \tilde{C}V^{-2}\, .
\end{equation}
In \cite{Hollands:2019}, $\tilde{C}$ was evaluated for the conformally coupled scalar field for a small number of spacetime parameters and was indeed found to be non-zero. Since one would expect $\tilde{C}$ to be an analytic function of the spacetime parameters, 
this strongly suggests that $\tilde{C}$ should be non-zero except for a measure zero set of parameters. Thus, sCC should be violated generically.

The analysis of \cite{Hollands:2019} combined functional analytic methods (for estimates) with semi-exact methods (for the value of $\tilde{C}$). However, 
the semi-exact method contained a numerical component, and the approach presented in \cite{Hollands:2019} was not really suitable for 
exploring a large range of spacetime parameters, and also not for generic masses of the scalar field \eqref{eq:kge}.  In this paper, we point out a variant of 
the method \cite{Hollands:2019} which seems much more efficient. The bottleneck of the calculation is to solve the radial wave equation semi-analytically to obtain the data for various scattering problems associated with RNdS. But rather than using series of hypergeometric functions and numerical integrations as in \cite{Hollands:2019}, we observe in this paper that the scattering data can also be obtained by local power series solutions. This brings a considerable amount of simplification for the remaining numerical analysis, enabling us to scan a wider parameter range. In the next section, we introduce the prerequisite notation and summarize some results by \cite{Hollands:2019} needed in the sequel. Section \ref{sec:ana} contains the analytic reformulation of the equations for the radial mode solutions of (\ref{eq:kge}). In section \ref{sec:num}, the results of the numerical computation are presented. We conclude in section \ref{sec:concl}.

\section{Setup}
\label{sec:setup}
In this work, we consider the RNdS spacetime, as shown in figure~\ref{fig:RNdS}. The metric is 
\begin{subequations}
\label{eq:st}
\begin{align}
\label{eq:metric}
g&=-f(r) \mathrm{d}t^2+f^{-1}(r)\mathrm{d}r^2+r^2\mathrm{d}\Omega^2\, , \\
\label{eq:fr}
f(r)&=-\frac{\Lambda}{3}r^2+1-\frac{2M}{r}+\frac{Q^2}{r^2}\, .
\end{align}
\end{subequations}
The parameters of the spacetime, $\Lambda>0$, $M>0$, and $Q$, represent the cosmological constant, the mass and the charge of the black hole. They are chosen in a physical region of parameter space. Then $f(r)$ has three real positive roots $r_-<r_+<r_c$, and one negative root $r_o=-(r_-+r_++r_c)<0$. The positive roots represent the Cauchy $(r_-)$, event $(r_+)$, and cosmological $(r_c)$ horizon. We will mostly be interested in the regime $M^2\approx Q^2$, where sCC is violated classically according to \cite{Cardoso:2017}. 

The surface gravity on the horizons is given by
\begin{equation}
\kappa_i=\frac{1}{2}|\partial_rf(r)|_{r=r_i}\, , \, \, \, \, \, \,  i\in \{-\,, \, \, +\, ,\, \, c\}\, .
\end{equation}

In the calculation we will use the tortoise coordinate $r_*$, which is defined by $\mathrm{d}r_*=f^{-1}(r)\mathrm{d}r$. More explicitly, we will fix the integration constants such that
\begin{align}
 \label{eq:rstar}
 r_*(r)&=\frac{1}{2\kappa_+}\ln |r_+-r| -\frac{1}{2\kappa_-}\ln |r_--r|\\ \nonumber &-\frac{1}{2\kappa_c}\ln |r_c-r| +\frac{1}{2\kappa_o}\ln |r_o-r|
 \end{align}
on all of the spacetime, in agreement with \cite{Hollands:2019}, eq. (12). In addition, on part $\rI$ of the spacetime, we define
the radial null coordinates 
\begin{align}
v &\equiv t+r_*\, , & u &\equiv t-r_*\, ,
\end{align}
where $u$ diverges to $+\infty$ on $\mathcal{H}^R$, while $v$ diverges to $+\infty$ on $\mathcal{H}_c^L$. These coordinates, and the metric, can be smoothly extended, by defining the Kruskal coordinates
\begin{align}
U &\equiv -e^{-\kappa_+u}\, , & V &\equiv -e^{-\kappa_-v}\, , 
\end{align}
of which $V$ can be smoothly extended across $\mathcal{C}\mathcal{H}^R$.

Using the relation $\partial v/ \partial V = \kappa_-^{-1} V^{-1}$ close to $\mathcal{C}\mathcal{H}^R$, it becomes clear, that (\ref{eq:TVVdiv}) holds if 
\begin{equation}
\langle T_{vv}(U,v)\rangle_\rU-\langle T_{vv}(U,v)\rangle_\rC \sim  \tilde{C}
\end{equation}
close to the Cauchy horizon.

In \cite{Hollands:2019}, it is explained how the constant $\tilde{C}$ can be calculated in terms of mode solutions to (\ref{eq:kge}) with boundary conditions given on $\mathcal{H}=\mathcal{H}^L\cup \mathcal{H}^-$ and $\mathcal{H}_c=\mathcal{H}_c^-\cup \mathcal{H}_c^R$. Here, we only give a brief summary of these results. The solutions used in the calculation for $\tilde{C}$ are two families of the so-called Boulware mode solutions. These solutions can be factorized as\footnote{Note that in \cite{Hollands:2019} the normalization of the modes was incorrect, leading also to an incorrect factor in the expression for $\tilde C$, which is corrected in (\ref{eq:finform}) below.}
\begin{equation}
\label{eq:psifact}
 \psi^{N}_{\omega \ell m}=(4\pi |\omega |)^{-\tfrac{1}{2}}Y_{\ell m}(\theta,\, \phi )e^{-i\omega t}F^{N}_{\omega\ell}(r)\, ,
\end{equation}
where $N$ is either $\rI$ or $\rII$. $Y_{\ell m}(\theta , \phi )$ are the spherical harmonics. $F^{N}_{\omega\ell}$ may be written as $r^{-1}R_{\omega \ell}^{N}(r)$, where $R_{\omega \ell}^{N}$ satisfy the differential equation
\begin{equation}
\label{eq:Rwl}
\left[\partial_{r_*}^2-V_\ell(r)+\omega^2\right]R_{\omega \ell}^N(r_*)=0\, ,
\end{equation}
with the smooth effective potential
\begin{equation}
V_{\ell}(r)=f(r)\left(\frac{\ell (\ell +1)}{r^2}+\frac{\partial_r f(r)}{r}+\mu^2\right)
\end{equation}
falling off faster than any power in $r_*$ towards each of the horizons.
As a result, the initial conditions on $\psi^{N}_{\omega \ell m}$ can be given in terms of the asymptotic behavior of $R_{\omega \ell}^{N}$ for large values of $|r_*|$:
\begin{subequations}
\label{eq:BoundaryCond}
\begin{align}
\label{eq:RupI}
R_{\omega \ell}^{\rI}(r)&=\begin{cases}
e^{i\omega r_*}+\mathcal{R}^{\rI}_{\omega \ell}e^{-i\omega r_*} & r_*\to -\infty\\
\mathcal{T}_{\omega \ell}^{\rI}e^{i\omega r_*} & r_*\to \infty\\
\end{cases}\\
\label{eq:RupII}
R_{\omega \ell}^{\rII}(r)&=\begin{cases}
e^{i\omega r_*} & r_*\to -\infty\\
\mathcal{T}_{\omega \ell}^{\rII}e^{i\omega r_*}+\mathcal{R}_{\omega \ell}^{\rII}e^{-i\omega r_*} & r\to \infty\, .\\
\end{cases}
\end{align}
\end{subequations}
In addition, since the Wronskian for the solutions to (\ref{eq:Rwl}) is constant, we find a relation between the coefficients $\mathcal{T}^N_{\omega\ell}$ and $\mathcal{R}^N_{\omega\ell}$:
\begin{subequations}
\begin{align}
\label{eq:TRcond1}
|\mathcal{R}_{\omega \ell}^{\rI}|^2+|\mathcal{T}_{\omega \ell}^{\rI}|^2=1\, , \\
\label{eq:TRcond2}
|\mathcal{T}_{\omega \ell}^{\rII}|^2-|\mathcal{R}_{\omega \ell}^{\rII}|^2=1\, .
\end{align}
\end{subequations}
$\mathcal{T}^N_{\omega\ell}$ and $\mathcal{R}^N_{\omega\ell}$ can then be used to determine the constant $\tilde{C}$. In particular, once the scattering coefficients are known, $\tilde{C}$ is given by
\begin{subequations}
\label{eq:Ctilde}
\begin{align}
\label{eq:finform}
\tilde{C}&=\sum_{\ell}\frac{2\ell +1}{16\pi^2 r_-^2}\int_{0}^{\infty}\mathrm{d}\omega \omega n_{\ell}(\omega )\, , \\
\label{eq:integrand}
n_\ell (\omega )&=\left| \mathcal{T}^{\rI}_{\omega \ell}\right|^2\left|\mathcal{T}^{\rII}_{\omega \ell}\right|^2\coth \frac{\pi \omega}{\kappa_c}\\ \nonumber 
&+\left(\left| \mathcal{R}^{\rI}_{\omega \ell}\right|^2\left| \mathcal{T}^{\rII}_{\omega \ell}\right|^2+\left|\mathcal{R}^{\rII}_{\omega \ell}\right|^2\right)\coth\frac{\pi\omega}{\kappa_+}\\ \nonumber
&+2\mathrm{csch} \frac{\pi\omega}{\kappa_+}\mathrm{Re}\left(\mathcal{R}^{\rI}_{\omega \ell}\overline{\mathcal{T}^{\rII}_{\omega \ell}}\,\overline{\mathcal{R}^{\rII}_{\omega \ell}}\right)-\coth \frac{\pi\omega}{\kappa_-} \, .
\end{align}
\end{subequations}

Thus, in order to determine the divergence of the energy flux near the Cauchy horizon, we need to evaluate the scattering coefficients for the radial part of the Boulware mode solutions.

\section{Analytic reformulation of the radial equation}
\label{sec:ana}
In this section, we will reformulate (\ref{eq:kge}) in order to find solutions of its radial part. We normalize these solutions such that they behave as $e^{\pm i\omega r_*} r_i^{-1}$ at the three horizons $i=+,\, -,\, c$.
In order to obtain these solutions, we employ a mode-type ansatz as in (\ref{eq:psifact}). The equation for its radial part  $F(r)_{\omega\ell}^{N}$, dropping the indices for convenience, then reads
\begin{equation}
\left[\partial_r\left(\Delta\partial_r\right) - \ell (\ell +1)+\frac{(\omega r^2 )^2}{\Delta}-\mu^2r^2\right]F(r)=0 \, ,
\end{equation}
with $\Delta=r^2f(r)$. As was demonstrated in \cite{Suzuki:1999}, this equation can be brought into a more manageable form by changing to the dimensionless variable 
\begin{equation}
x=\frac{r_--r_o}{r_--r_+}\frac{r-r_+}{r-r_o}\equiv x_{\infty}\frac{r-r_+}{r-r_o}\, .
\end{equation}
In the coordinate $x$, the event horizon is located at $x=0$, while the Cauchy horizon is at $x=1$.
We then define the constants
\begin{subequations}
\begin{align}
 a_+&=\frac{i\omega}{2\kappa_+} \, , & a_-&=-\frac{i\omega}{2\kappa_-} \, , \\
 a_c&=-\frac{i\omega}{2\kappa_c} \, , & a_o&=\frac{i\omega}{2\kappa_o} \, .
 \end{align}
 \end{subequations}
which can be shown to satisfy \cite{Suzuki:1999}
 \begin{equation}
 \label{eq:arel}
 a_++a_-+a_c+a_o=0\, .
\end{equation}
After these definitions, we can now write down an ansatz for $F(x)$:
 \begin{equation}
 F(x)=|x|^{a_+}|1-x|^{a_-}\left| \frac{x-x_c}{1-x_c}\right|^{a_c}\left(\frac{x-x_{\infty}}{1-x_{\infty}}\right) h(x)\, ,
 \end{equation}
 where $x_c=x(r_c)$. This choice of signs for the exponents allows us to interpret the prefactor of $h(x)$ in terms of the tortoise coordinate $r_*$. With the explicit form for $r_*$ given by (\ref{eq:rstar}), it can be rewritten as 
\begin{align}
\label{eq:prefac}
&|x|^{a_+}|1-x|^{a_-}\left| \frac{x-x_c}{1-x_c}\right|^{a_c}\left(\frac{x-x_{\infty}}{1-x_{\infty}}\right)\\ \nonumber
&= e^{i\omega r_*}e^{i\omega D}\frac{x_{\infty}}{x_{\infty}-1}\frac{r_+-r_o}{r-r_o}\, .
\end{align} 
The last factor is a bounded function for $r>0$, and $D$ is a constant depending only on the parameters of the spacetime.
The identification (\ref{eq:prefac}) allows us to construct solutions with the desired free wave behavior near the horizons, once a solution for $h(x)$ which is regular at the corresponding horizon, is determined.

The equation for $h(x)$ reads
\begin{widetext}
 \begin{equation}
 \label{eq:fx}
\partial_x^2h(x)+\left[\frac{\gamma}{x}+\frac{\delta}{x-1}+\frac{\epsilon}{x-x_c}\right]\partial_xh(x)+
\left[\frac{\sigma_+\sigma_-x-q}{x(x-1)(x-x_c)}+\frac{\Delta_1x-\Delta_2}{x(x-1)(x-x_c)(x-x_{\infty})^2}\right]h(x)=0\, .
\end{equation}
\end{widetext}
 The constants appearing in the second term on the left-hand side are given by 
 \begin{align}
\gamma=1+2a_+ \, ,  & & \delta=1+2a_- \, ,& &\epsilon=1+2a_c \, .
\end{align}
They are connected to 
\begin{subequations}
\begin{align}
\sigma_+= 1+a_++a_-+a_c-a_o&=1-2a_o\\
\sigma_-=1+a_++a_-+a_c+a_o&=1
\end{align}
\end{subequations}
by the relation
\begin{equation}
\gamma +\delta +\epsilon = \sigma_+ + \sigma_- +1\, .
\end{equation}
The parameter $q$ is related to the parameter $v$ given in \cite{Suzuki:1999} by $q=-v(a=s=e=0)$, with a slight change due to the mass of the scalar field, and reads, in our case,
\begin{align}
q&=\frac{-6\omega^2r_+^3\left(r_+r_--2r_-r_c+r_+r_c\right)}{\Lambda(r_+-r_-)^3(r_+-r_c)^2(r_+-r_o)(r_c-r_o)}\nonumber \\ 
&+x_{\infty}+\left[(1+x_c)a_++x_ca_-+a_c\right]\\ \nonumber
&+2a_+\left[x_ca_-+a_c\right]+\frac{3\ell ( \ell +1)}{\Lambda (r_+-r_-)(r_c-r_o)}\\ \nonumber
 &+\frac{3\mu^2r_o^2}{\Lambda (r_+-r_-)(r_c-r_o)}\, .
\end{align}
The constants in the last term in (\ref{eq:fx}) are given by
\begin{subequations}
\begin{align}
\Delta_1&=\left(2-\frac{3\mu^2}{\Lambda}\right)\frac{2r_o(r_o-r_+)}{(r_+-r_-)(r_c-r_o)}\\
\Delta_2&=\left(2-\frac{3\mu^2}{\Lambda}\right) \frac{x_{\infty}^2\left(r_o^2-r_+^2\right)}{(r_+-r_-)(r_c-r_o)}\,.
\end{align}
\end{subequations}
Note that $\Delta_1$ and $\Delta_2$ are both zero in the case of a conformally coupled scalar, $\mu^2=\tfrac{2}{3}\Lambda$. In this case, the equation for $h(x)$ reduces to a Heun equation \cite{Ronveaux:1995}. The  method of \cite{Suzuki:1999}, which was also used in \cite{Hollands:2019}, works only in this conformally coupled case. Instead, we here employ a method analogous to that described in \cite{Ronveaux:1995}, ch.A3, to solve the equation in the case of general $\mu^2$. In order to do so, we first consider $h(x)$ in a neighborhood of $x=0$, which corresponds to a neighborhood of the event horizon $r=r_+$, and make a power-series ansatz for $h(x)$: 
\begin{equation}
\label{eq:powerseries}
h_+(x)= \sum\limits_{n=0}^{\infty}h_n x^n \, .
\end{equation}
We take only the one-sided power series in order to obtain a regular solution at $x=0$. Moreover, we normalize the function at $x=0$ by setting $h_0=1$. (\ref{eq:fx}) then yields a five-term recurrence relation for $h_n$,
\begin{align}
\label{eq:recrel}
&x_{\infty}^2a(n+2)h_{n+2}\\\nonumber
&-[x_{\infty}^2 b(n+1)+2x_{\infty}a(n+1)+\Delta_2]h_{n+1}\\ \nonumber
&+[x_{\infty}^2 c(n)+2x_{\infty} b(n)+ a(n)+\Delta_1]h_n\\ \nonumber
&-[2x_{\infty}c(n-1)+b(n-1)]h_{n-1}\\ \nonumber
&+c(n-2)h_{n-2}=0 \, ,
\end{align}
where
\begin{subequations}
\begin{align}
a(n)&=x_c n(n-1+\gamma )\\
b(n)&=n\left[ (x_c+1)(n-1+\gamma )+x_c \delta +\epsilon\right]+q\\
c(n)&=(n+\sigma_+)(n+\sigma_-)\, .
\end{align}
\end{subequations}
Note that in the conformally coupled case, (\ref{eq:recrel}) reads
\begin{align}
&x_{\infty}^2[a(n+2)h_{n+2}- b(n+1)h_{n+1}+c(n)]h_n]\\\nonumber
&-2x_{\infty}[a(n+1)h_{n+1}- b(n)h_n+c(n-1)h_{n-1}]\\ \nonumber
&+a(n)h_n-b(n-1)h_{n-1}+c(n-2)h_{n-2}=0 \, .
\end{align}
The left-hand side vanishes if the $h_n$ satisfy 
\begin{align}
a(n+1)h_{n+1}- b(n)h_n+c(n-1)h_{n-1}=0\, ,
\end{align}
reducing the five-term to a three-term recurrence relation in this case.

\begin{figure}
\includegraphics[width=0.45\textwidth]{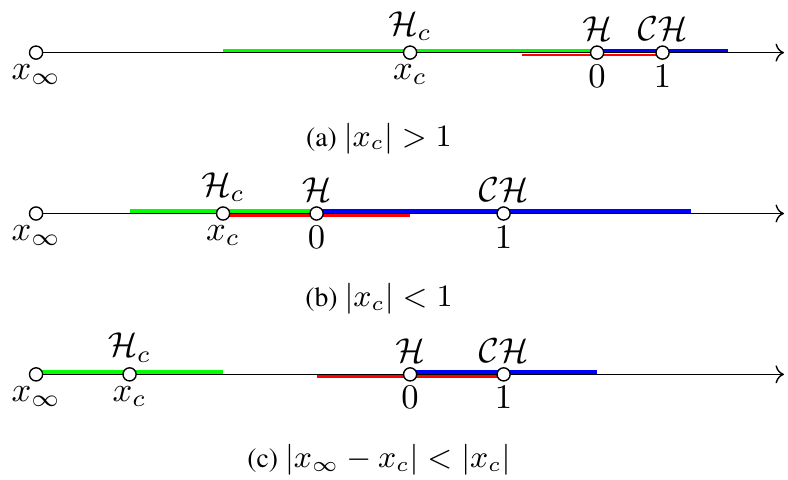}
\caption{\footnotesize The correspondence between the three singular points $0$, $1$ and $x_c$ of the equation (\ref{eq:fx}) for $h(x)$ and the three horizons. The first two  cases correspond to $|x_c|>1$ and $|x_c|<1$. The coloured regions indicate the region of convergence for the three solutions $h_+(x)$ (red), $h_-(x)$ (blue), and $h_c(x)$ (green). The third picture demonstrates what happens if $|x_{\infty}-x_c|<|x_c|$: the solutions $h_c(x)$ and $h_+(x)$ have no overlap, which would be fatal for our computation. However, this case does not occur for physical spacetime parameters.}
\label{fig:roc}
\end{figure}

The initial conditions for the recurrence relation are given by $h_n=0$ for $n<0$ and $h_0=1$. They uniquely determine all $h_n$, and thus $h_+(x)$. In order to investigate the radius of convergence for $h_+(x)$, we divide (\ref{eq:recrel}) by $n^2$ and $h_{n-2}$, and define $\rho_n=\tfrac{h_{n+1}}{h_n}$.
Then, assuming that $\lim_{n\to \infty}\rho_n=\rho$, we get in the limit of large $n$,
\begin{align}
0&=x_{\infty}^2x_c\rho^4-\left[x_{\infty}^2(x_c+1)+2x_{\infty}x_c\right]\rho^3\\ \nonumber
&+\left[ x_{\infty}^2+2x_{\infty}(x_c+1)+x_c\right]\rho^2\\ \nonumber
&-\left[ 2x_{\infty}+x_c+1\right]\rho+1\\\nonumber
&=(1-x_{\infty}\rho)^2(1-\rho)(1-x_c\rho )
\end{align}
Thus, $\rho \in \left\{\tfrac{1}{x_{\infty}} , \, \tfrac{1}{x_c} , \, 1\right\}$ and the radius of convergence for $h_+(x)$ will in general be 
\begin{align}
R=\min\{1,\, |x_c|,\, |x_{\infty}| \}\, .
\end{align}
In other words, $h_+(x)$ is defined in a region around the event horizon $r_+$, in both directions, $r<r_+$ and $r>r_+$, up to the closest other horizon in either direction, in terms of $x$. This is indicated in figure (\ref{fig:roc}) by the red interval. In order to cover also the Cauchy horizon and the cosmological horizon, we need two other solutions for $h(x)$, expanded around $x=1$ and $x=x_c$.

This can be achieved by using the coordinate change transformations of the Heun equation \cite{Ronveaux:1995}. We find that, even with the additional terms $\Delta_1$ and $\Delta_2$, which would be zero for a Heun equation, these transformations yield a differential equation for $h(x)$ of the same form as (\ref{eq:fx}). In these new coordinates $z$, one of the two singular points $x=1$ or $x=x_c$ will now be located at $z=0$, and the constants appearing in (\ref{eq:fx}) need to be transformed as well. Once this is done, the solution of $h(z)$ around $z=0$ will then be constructed completely analogously to $h_+(x)$. 

In particular, for the solution near the Cauchy horizon, $x=1$, let $\tilde{x}=1-x$. Then the new constants are
\begin{subequations}
\begin{align}
\tilde{\gamma}&=\delta \, ,  & \tilde{\delta}&=\gamma \, ,  \\ 
\tilde{\epsilon}&=\epsilon \, , & \tilde{\sigma}_+&= \sigma_+ \, ,  \\ 
\tilde{\sigma}_-&=\sigma_- \, ,  & \tilde{q}&=\sigma_+\sigma_--q\, , \\ 
\tilde{\Delta}_1&=\Delta_1\, ,  & \tilde{\Delta}_2 &= \Delta_1-\Delta_2\, ,\\ 
\tilde{x}_c&=1-x_c\, ,  & \tilde{x}_{\infty}&=1-x_{\infty}\, .
\end{align}
\end{subequations}
The solution $h_-(x)$ around the Cauchy horizon is then a power series in $1-x$. The coefficients for that series are determined by the recurrence relation (\ref{eq:recrel}), where in the definition of $a(n)$, $b(n)$ and $c(n)$, all constants are replaced by the transformed ones given above. The radius of convergence for the solution $h_-(x)$ around $x=1$ is then $\Tilde{R}=\min\{ 1, |\tilde{x}_c|, \, |\tilde{x}_{\infty}|\}$. It is indicated in figure \ref{fig:roc} by the blue region.

For the solution around the cosmological horizon, $x=x_c$, choose $x^{\prime}=\frac{x_c-x}{x_c-1}$. Under this coordinate transformation, the constants in (\ref{eq:fx}) change as 
\begin{subequations}
\begin{align}
\gamma^{\prime}&=\epsilon\, , &  \delta^{\prime}&=\delta \, , \\
\epsilon^{\prime}&=\gamma \, ,& \sigma_+^{\prime}&=\sigma_+ \, , \\ 
\sigma_-^{\prime}&=\sigma_-\, , & q^{\prime}&=\frac{x_c\sigma_+\sigma_--q}{x_c-1}\, , \\ 
\Delta_1^{\prime}&=\frac{\Delta_1}{(x_c-1)^2}\, , & \Delta_2^{\prime}&=\frac{x_c\Delta_1-\Delta_2}{(x_c-1)^3}\, , \\ 
x_c^{\prime}&=\frac{x_c}{x_c-1} \, , &  x_{\infty}^{\prime}&=\frac{x_c-x_{\infty}}{x_c-1}\, .
\end{align}
\end{subequations}
The third solution $h_c(x)$ will then be a power series in $(x_c-x)/(x_c-1)$, with coefficients again given by the recurrence relation (\ref{eq:recrel}), utilizing the primed constants.
The radius of convergence for $h_c(x)$ in terms of $x$ around $x_c$ is given by $R^{\prime}=\min\{|1-x_c|, |x_c|, |x_{\infty}-x_c|\}$. It is indicated in figure~\ref{fig:roc} by the green line.

Note that in neither case is the radius of convergence given by the third option in the brackets, which depends on $x_{\infty}$. The reason is that not only $|x_{\infty}|>1$ for all physical choices of the spacetime parameters, but also
\begin{align}
&|x_{\infty}-x_c|-|x_c|\\ \nonumber
&=|x_{\infty}|\left(\left|1-\frac{r_c-r_+}{r_c-r_o}\right|-\frac{r_c-r_+}{r_c-r_o}\right) \\ \nonumber
&=|x_{\infty}|\left(\frac{r_c-r_o-2r_c+2r_+}{r_c-r_o}\right)\\ \nonumber
&=|x_{\infty}| \left(\frac{3r_++r_-}{r_c-r_o}\right)>0\, ,
\end{align}
where we have used the definition of $r_o$.
 This is especially important for $h_c(x)$. If $R^{\prime}$ were determined by $|x_\infty-x_c|$, the region of convergence of the solution $h_c(x)$ around the cosmological horizon would not overlap with the region of convergence of $h_+(x)$, as in the example in figure~\ref{fig:roc}c. This overlap, however, is crucial in order to determine the scattering coefficients. That is also the reason why this formalism does not work very well when one is close to extremality, $r_+-r_-\lesssim 0.01 (r_c-r_+)$. In this case, $|x_c|$ is very large compared to $1$. As a result, $h_+(x)$ overlaps with $h_c(x)$ in a region where the convergence of the power series in $h_c(x)$ is already very slow. This makes the function hard to compute numerically, and also increases numerical errors. While this does not make the computation by our method impossible in principle in this regime, other methods might be more suitable. For this reason we will mostly restrict to the parameter region $Q\leq M$.

Now, we have three solutions of the radial wave equation, $F_-(x)$, $F_+(x)$ and $F_c(x)$, each of them valid in a neighborhood of one of the horizons, by inserting the three solutions for $h(x)$ into the ansatz for $F(x)$. We then define the three normalized solutions to the radial equation as
\begin{subequations}
\begin{align}
R_{\omega \ell}^-&=e^{-i\omega D}r_-^{-1}F_-(x)\, , \\
R_{\omega \ell}^+&=e^{-i\omega D}\left(\frac{x_{\infty} -1}{x_{\infty}}\right)r_+^{-1}F_+(x)\, , \\
R_{\omega \ell}^c&=e^{-i\omega D}\left(\frac{1-x_{\infty}}{x_c-x_{\infty}}\right)r_c^{-1}F_c(x)\, .
\end{align}
\end{subequations}
The radial mode functions $r^{-1}R_{\omega \ell}^{N}(r)$, for $N\in\{\rI , \rII\}$, of the Boulware mode solutions can now be constructed out of the normalized solutions given above around each of the three horizons. Consider first $R_{\omega \ell}^{\rI}(r)$, whose asymptotic behavior is described in (\ref{eq:RupI}). We can determine the scattering coefficients by matching the two expressions for this functions, obtained from  $R_{\omega \ell}^+$ and $R_{\omega \ell}^c$, and their first derivatives, 
\begin{subequations}
\label{eq:RTeqI}
\begin{align}
\overline{R_{\omega \ell}^+}(x)+\mathcal{R}_{\omega \ell}^{\rI}R_{\omega \ell}^+(x)&=\mathcal{T}_{\omega \ell}^{\rI}\overline{R_{\omega \ell}^c}(x)\, , \\ 
\partial_x\overline{R_{\omega \ell}^+}(x)+\mathcal{R}_{\omega \ell}^{\rI}\partial_xR_{\omega \ell}^+(x)&=\mathcal{T}_{\omega \ell}^{\rI}\partial_x\overline{R_{\omega \ell}^c}(x)\, ,
\end{align}
\end{subequations}
at some $x$ in the overlap of the functions $R_{\omega \ell}^+$ and $R_{\omega \ell}^c$. Analogously, $\mathcal{R}_{\omega \ell}^{\rII}$ and $\mathcal{T}_{\omega \ell}^{\rII}$ are determined by solving
\begin{subequations}
\label{eq:RTeqII}
\begin{align}
R_{\omega \ell}^+(x)&=\mathcal{T}_{\omega \ell}^{\rII}R_{\omega \ell}^-(x)+\mathcal{R}_{\omega \ell}^{\rII}\overline{R_{\omega \ell}^-}(x)\, , \\ 
\partial_x R_{\omega \ell}^+(x)&=\mathcal{T}_{\omega \ell}^{\rII}\partial_x R_{\omega \ell}^-(x)+\mathcal{R}_{\omega \ell}^{\rII}\partial_x \overline{R_{\omega \ell}^-}(x)
\end{align}
\end{subequations}
for some $x$ in the overlap of $R_{\omega \ell}^+(x)$ and $R_{\omega \ell}^-(x)$.
\section{Numerical results}
\label{sec:num}
The numerical setup for this calculation is very straightforward. For a chosen set of parameters $\{r_-,\, r_+,\, r_c,\, \omega ,\, \ell , \, \mu^2\}$, the five-term recurrence relation (\ref{eq:recrel}) is calculated numerically up to some large $n$ for the three horizons. For better comparability with \cite{Cardoso:2017}, we will actually use the parameter set $\{Q,\, \Lambda,\,  \omega ,\, \ell , \, \mu^2\}$ and normalise everything with respect to the black hole mass $M$. The resulting approximations for $R_{\omega \ell}^+(x)$, $R_{\omega \ell}^-(x)$ and $R_{\omega \ell}^c(x)$ are then evaluated at some $-1<x<1$. We make a rough estimate of the cutoff error by the relative contribution of the last term at our evaluation points. It appears that $\sim$5000 terms are sufficient to keep this error below $\mathcal{O}(10^{-15})$. After that, (\ref{eq:RTeqI}) and (\ref{eq:RTeqII}) are solved numerically. The numerical precision for the calculation of the recurrence relation and for solving (\ref{eq:RTeqI}) and (\ref{eq:RTeqII}) are chosen such that the numerical errors are of the absolute size $10^{-40}$. Thus the dominant contribution to the errors of the scattering coefficients comes from cutting off the series. From the result for the transmission and reflection coefficient we can calculate the integrand of (\ref{eq:finform}), $\omega n_{\ell}(\omega )$. In order to estimate the integral over $\omega$ and the sum over $\ell$, the above steps are repeated for different $\ell$ and $\omega$.

There are some checks which can be performed on the results. Firstly, for some example parameters we confirm that if (\ref{eq:RTeqI}) holds, then at the same time the equation also holds for higher $x$-derivatives. We found the errors to be of the order expected from the cutoff error. Secondly, for some example parameters we compared our results to the ones from numerical integration of the radial differential equation. The results are in agreement up to the $0.02\%$-level for $\omega r_+$ around 1, the agreement being even better for smaller $\omega$. Thirdly, we verify that the transmission and reflection coefficients satisfy the relations (\ref{eq:TRcond1}) and (\ref{eq:TRcond2}) respectively.
The deviation from these relations also helps us to estimate the numerical error made by cutting off the recurrence relation after a finite number of recursions. We make sure that the deviation does not become larger than $\mathcal{O}(10^{-15})$. We find that this can also be achieved with 5000 terms of the recurrence relation for all our parameter settings.

 \begin{figure*}
\begin{subfigure}{0.4\textwidth}
\includegraphics[scale=0.45]{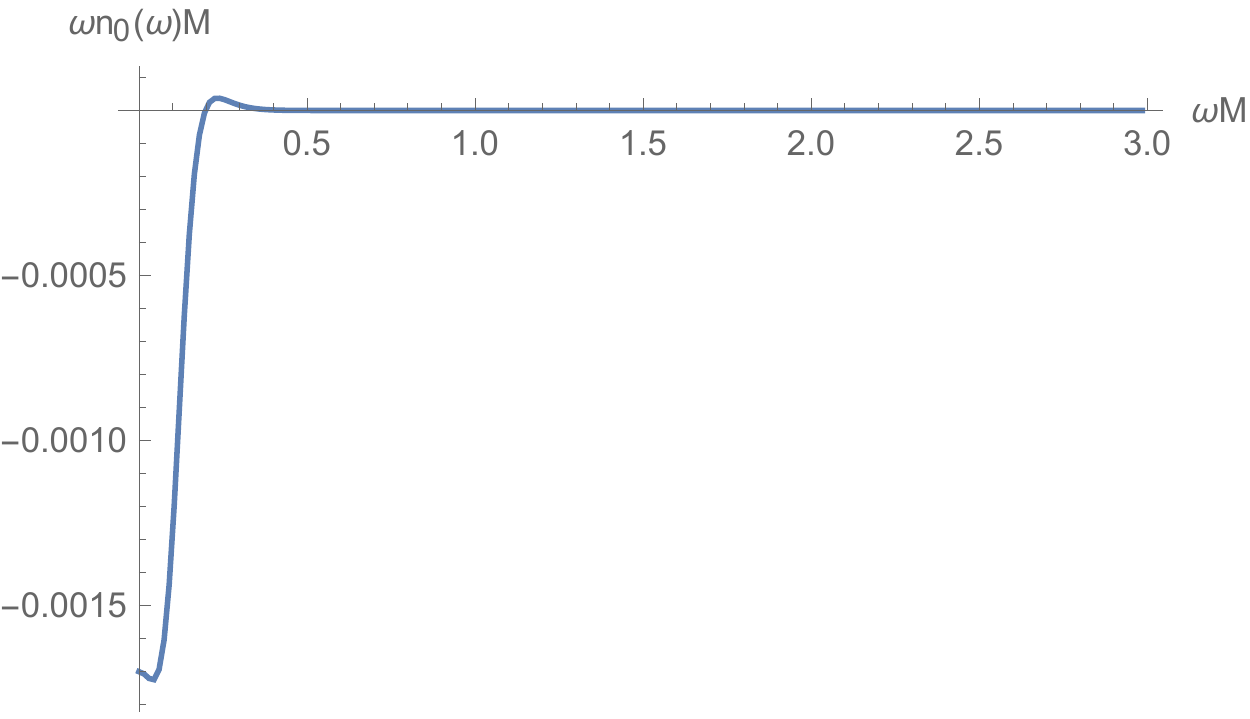}
\caption{\footnotesize $\omega n_0(\omega )M$, $\mu^2=0$, $\Lambda=0.02M^{-2}$ and $Q=0.9917M$}
\label{fig:Ldf1}
\end{subfigure}
\begin{subfigure}{0.4\textwidth}
\includegraphics[scale=0.45]{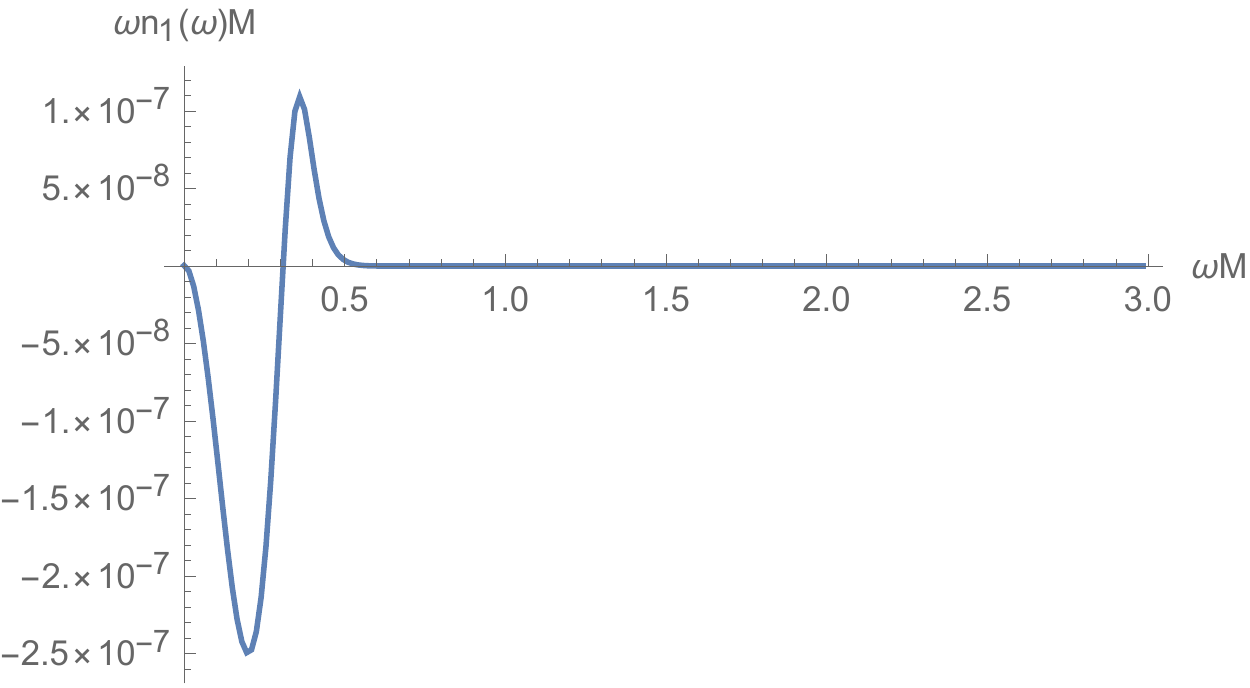}
\caption{\footnotesize$\omega n_1(\omega )M$, $\mu^2=0$, $\Lambda=0.02M^{-2}$ and $Q=0.9917M$}
\label{fig:Ldf2}
\end{subfigure}
\begin{subfigure}{0.4\textwidth}
\includegraphics[scale=0.45]{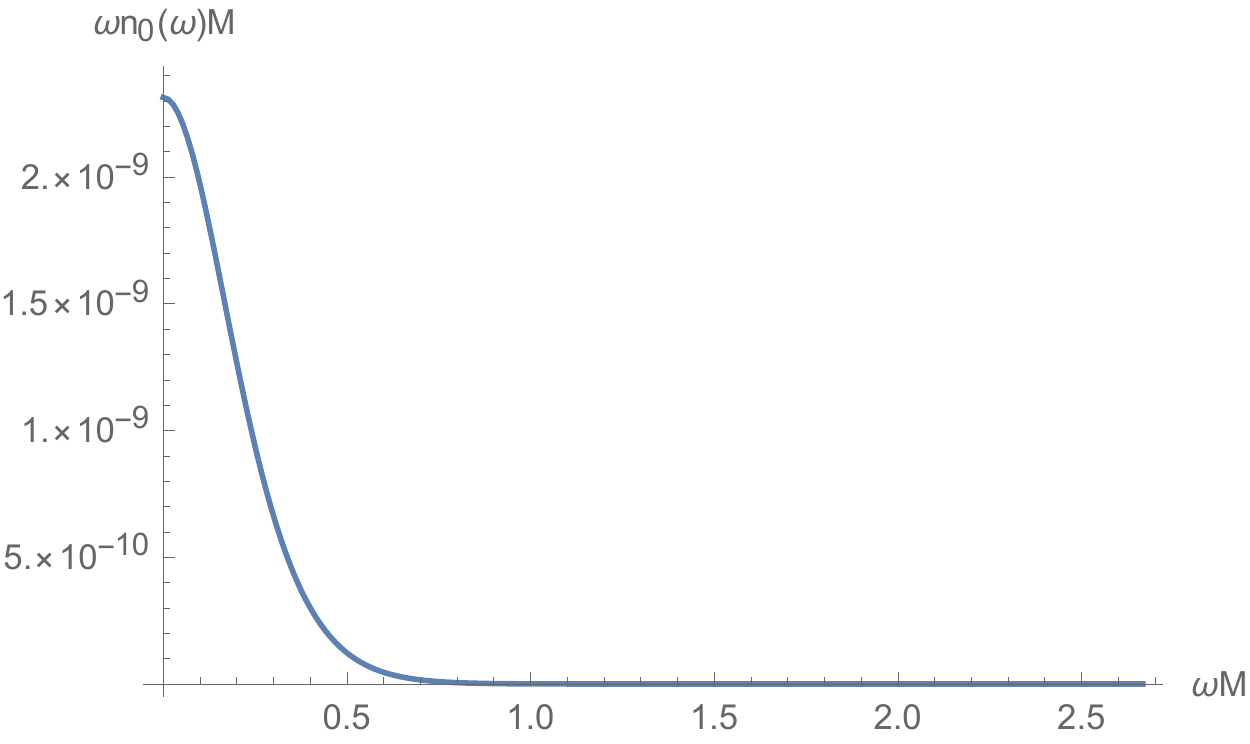}
\caption{\footnotesize$\omega n_0(\omega )M$, $\mu^2=1000/3 \Lambda$, $\Lambda=0.14M^{-2}$ and $Q=0.9945M$}
\label{fig:Lds1}
\end{subfigure}
\begin{subfigure}{0.4\textwidth}
\includegraphics[scale=0.45]{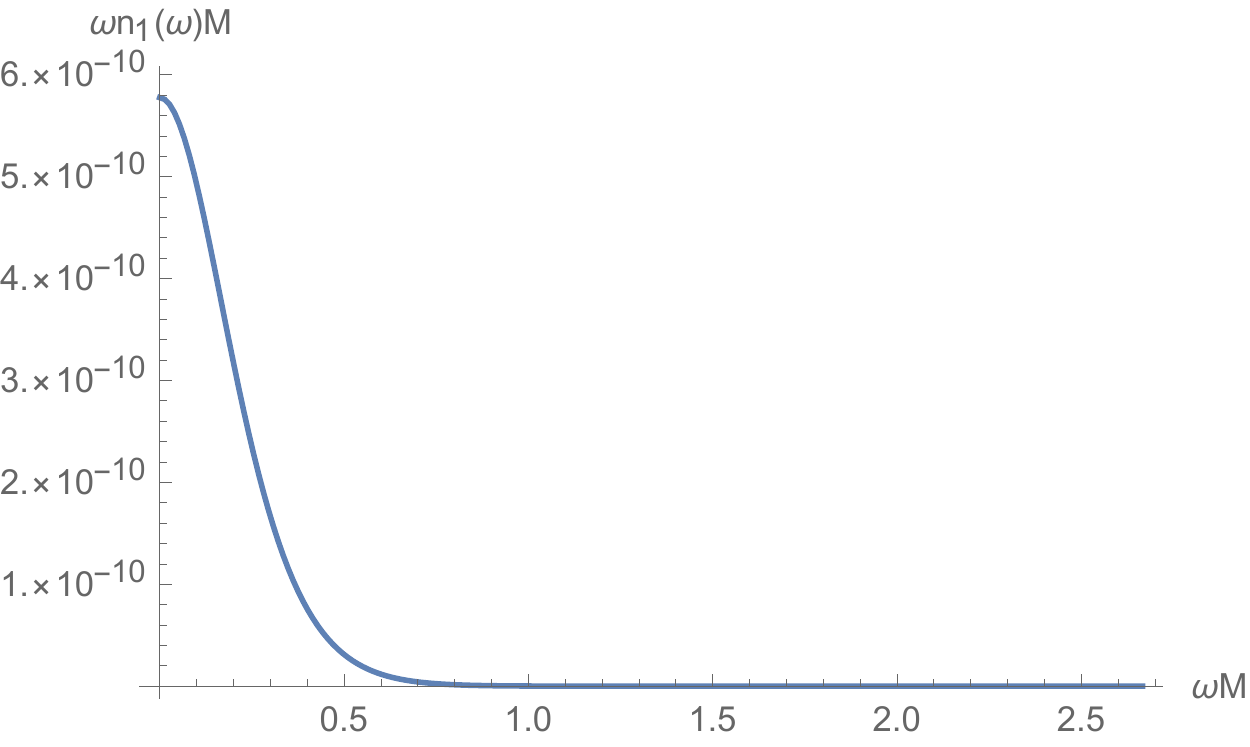}
\caption{\footnotesize$\omega n_1(\omega )M$, $\mu^2=1000/3 \Lambda$, $\Lambda=0.14M^{-2}$ and $Q=0.9945M$}
\label{fig:Lds2}
\end{subfigure}
\caption{\footnotesize The integrands $\omega n_0(\omega )M$ and $\omega n_1(\omega )M$ for a massless scalar with $\Lambda=0.02M^{-2}$ and $Q=0.9917M$ ((\ref{fig:Ldf1}) and (\ref{fig:Ldf2})), as well as for a scalar of mass $1000/3 \Lambda$ with $\Lambda=0.14M^{-2}$ and $Q=0.9945M$ ((\ref{fig:Lds1}) and (\ref{fig:Lds2})).}
\label{fig:Ldecay}
\end{figure*}
Concerning the convergence of the sum over $\ell$, our results indicate that the maximal value of the integrand $\omega n_{\ell}(\omega )$ always decreases by some factor $k>1$ when going from $\ell$ to $\ell+1$. It turns out that $k$ is larger for small values of $\Lambda$ and small values of the scalar field mass $\mu^2$. For the massless scalar and $\Lambda=0.02M^{-2}$, it is of the order of $10^4$. For large scalar field masses, $\mu^2\approx 50 M^{-2}$, it can become as small as $5-10$ or even smaller. An example of this behaviour is shown in figure~\ref{fig:Ldecay}. It displays the integrand for $\ell=0$ and $\ell=1$ for a massless scalar field in a spacetime with $\Lambda=0.02M^{-2}$ and $Q=0.9917M$, as well for a field of mass $\mu^2=1000/3\Lambda$ in a spacetime with $\Lambda=0.14M^{-2}$ and $Q=0.9945M$.  This indicates that the interchange of limits performed in \cite{Hollands:2019}, eq. (127), is indeed justified, as long as the scalar field mass and the cosmological constant are sufficiently small. Moreover, the series in $\ell$ seems to be well approximated by the first few terms, the exact number depending on the desired accuracy and the spacetime parameters. For our results, we consider all $\ell$ for which $\max\limits_{\omega} |\omega n_\ell(\omega)M|>10^{-15}$.

Finally, the integral over $\omega$ is estimated by the mean Riemann sum. The errors due to this discrete integral approximation are estimated by the upper and lower Riemann sums. We include values of $\omega$ up to some $\omega_{\text{max}}$, where the integrand $\omega n_{\ell}(\omega )$ has sufficiently decayed such that the rest of the integral can be neglected. Here, $\omega_{\text{max}}$ is chosen such that the integrand has decayed below $10^{-15}$. This value, which already appeared in the estimates for the cutoff errors, has been chosen such that the dominant contribution to the error comes from the approximation of the integral. We find that for most the parameters we tested, $\omega_{\text{max}}=3M^{-1}$ as a threshold seems sufficient, while for a few cases of comparably low black hole charge $Q$ we choose $\omega_{\text{max}}=4.5M^{-1}$. 

\begin{figure}
\includegraphics[scale=0.5]{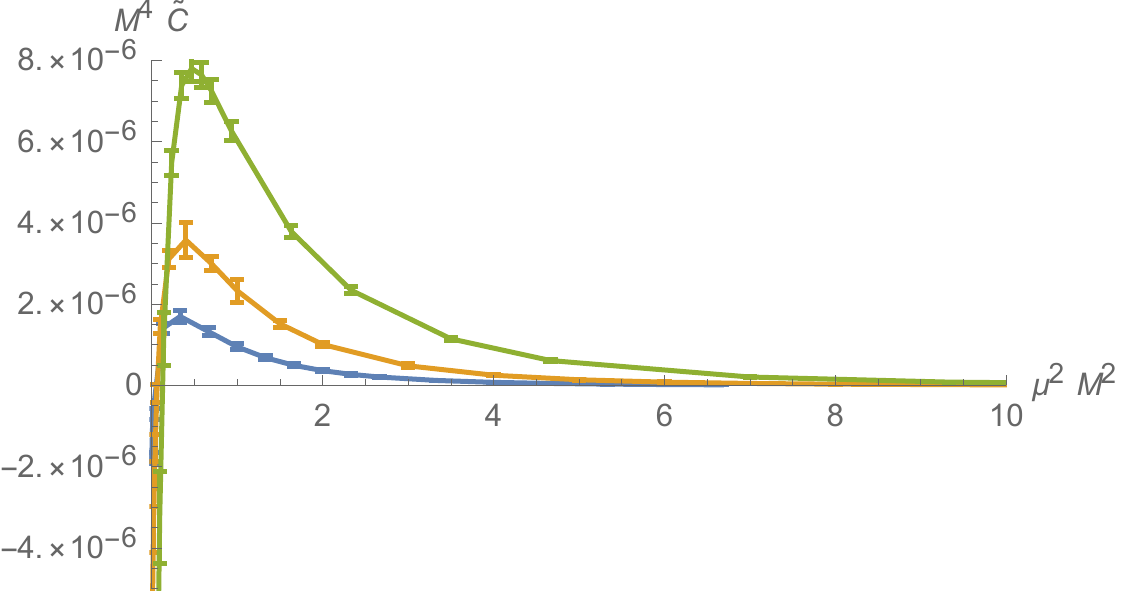}
\caption{\footnotesize$\langle T_{vv}\rangle \sim \tilde C$ at the Cauchy horizon in RNdS as a function of the mass $\mu^2$ of the scalar field. The spacetime parameters chosen are the ones which correspond approximately to the least critical black hole parameters with $\Lambda = 0.02M^{-2}$ (blue), $\Lambda=0.06M^{-2}$ (orange), and $\Lambda=0.14M^{-2}$ (green), such that sCC is classically violated \cite{Cardoso:2017}.}
\label{fig:CplotCard1}
\end{figure}

Figure~\ref{fig:CplotCard1} shows the parameter $\tilde{C}$ as a function of the mass $\mu^2$ of the scalar field in units of the black hole mass parameter $M$. Reinstating the gravitational constant $G$ explicitly, $\mu^2$ can be expressed in terms of more familiar units for particle masses as
\begin{align}
\mu^2=\left(\mu^2M^2\right)\frac{1.785\cdot 10^{-38}}{(M[M_\odot ])^2}\left(\frac{\text{GeV}}{\text{c}^2}\right)^2\, .
\end{align}
Here, $\mu^2M^2$ is the variable displayed on the x-axis in figure~\ref{fig:CplotCard1}.
Hence, the values of the scalar field mass considered here are still very small compared to the Higgs mass for example, at least for solar mass black holes.

\begin{figure*}
\begin{subfigure}{0.3\textwidth}
\includegraphics[scale=0.4]{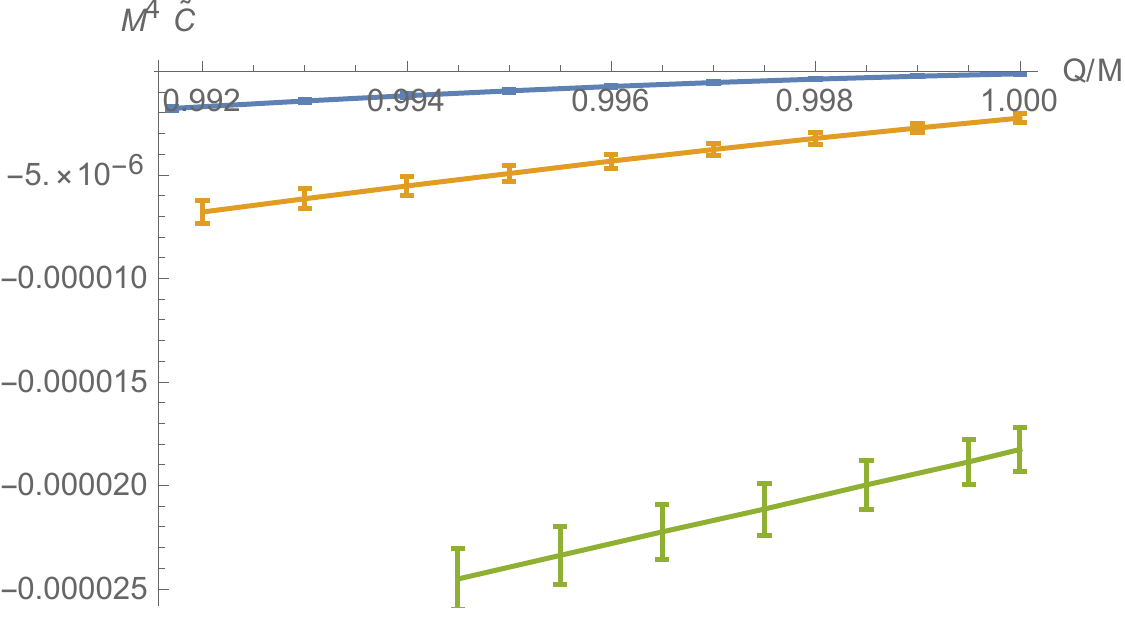}
\caption{\footnotesize$\langle T_{vv}\rangle \sim \tilde C$ at the Cauchy horizon in RNdS for a massless scalar field.}
\label{fig:CplotCard2}
\end{subfigure}
\begin{subfigure}{0.3\textwidth}
\includegraphics[scale=0.38]{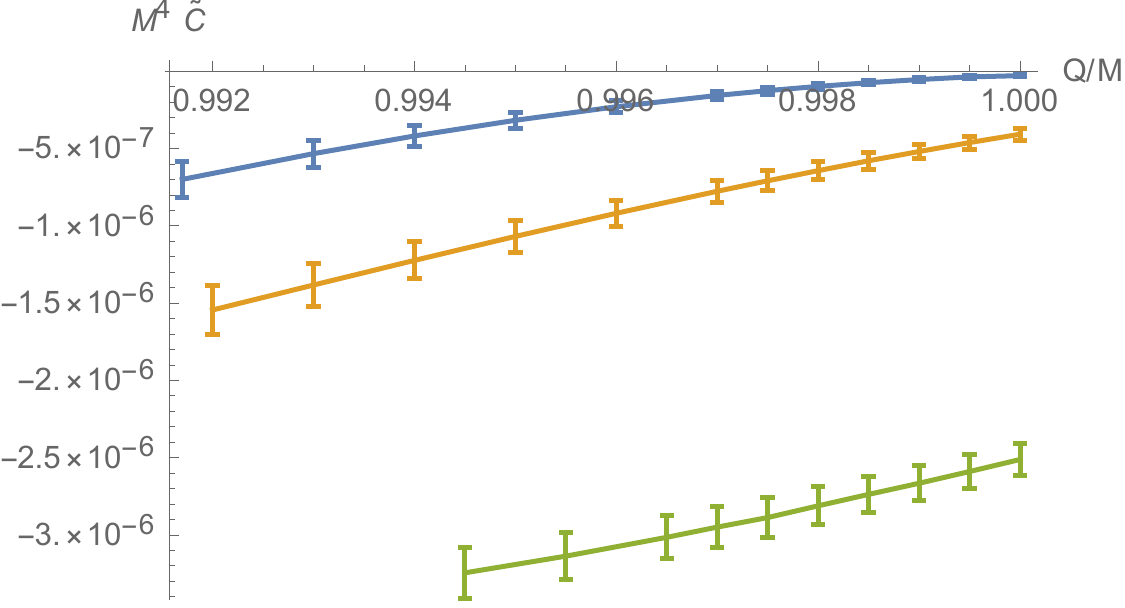}
\caption{\footnotesize$\langle T_{vv}\rangle \sim \tilde C$ at the Cauchy horizon in RNdS for a conformally coupled real scalar field.}
\label{fig:CplotCard3}
\end{subfigure}
\begin{subfigure}{0.3\textwidth}
\includegraphics[scale=0.38]{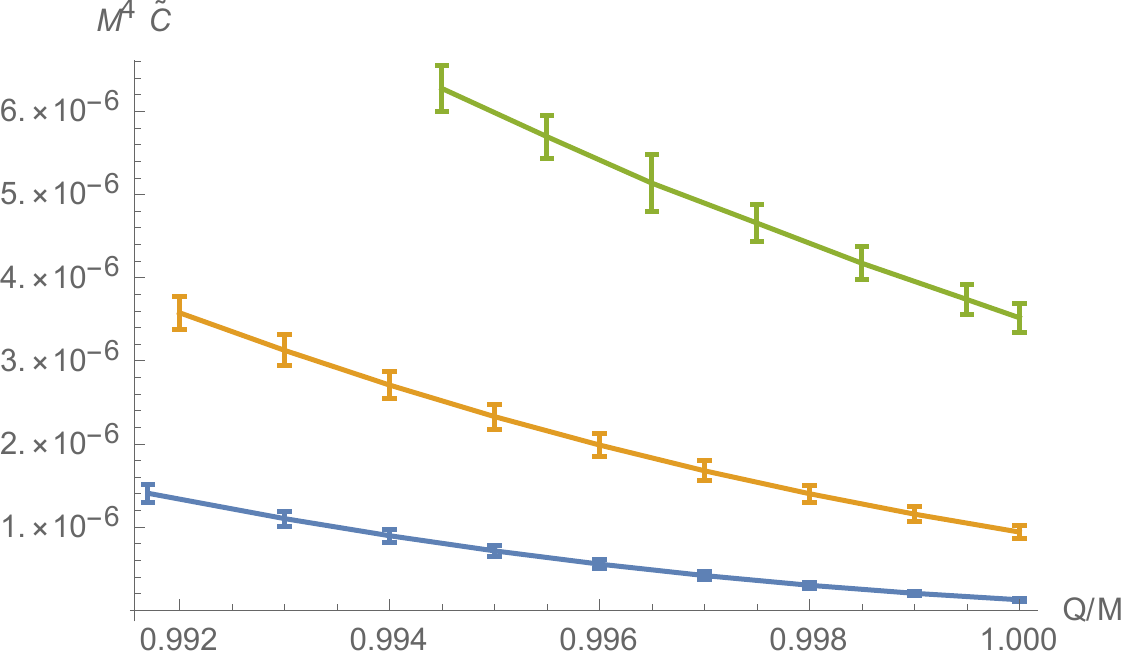}
\caption{\footnotesize$\langle T_{vv}\rangle \sim \tilde C$ at the Cauchy horizon in RNdS for real scalar field of mass $\mu^2=20/3\Lambda$.}
\label{fig:CplotCard4}
\end{subfigure}
\caption{\footnotesize$\langle T_{vv}\rangle \sim \tilde C$ at the Cauchy horizon in RNdS for a real scalar field. The three subplots show the massless (\ref{fig:CplotCard2}), conformally coupled (\ref{fig:CplotCard3}), and $\mu^2=20/3\Lambda$ (\ref{fig:CplotCard4}) case. The cosmological constant is fixed to $\Lambda=0.02M^{-2}$ (blue)/ $0.06M^{-2}$ (orange)/ $0.14M^{-2}$ (green). $\tilde{C}$ is plotted as a function of $Q/M$ in the part of the parameter regime, where sCC is violated classically \cite{Cardoso:2017} and $Q\leq M$.}
\end{figure*} 

\begin{figure}
\includegraphics[scale=0.5]{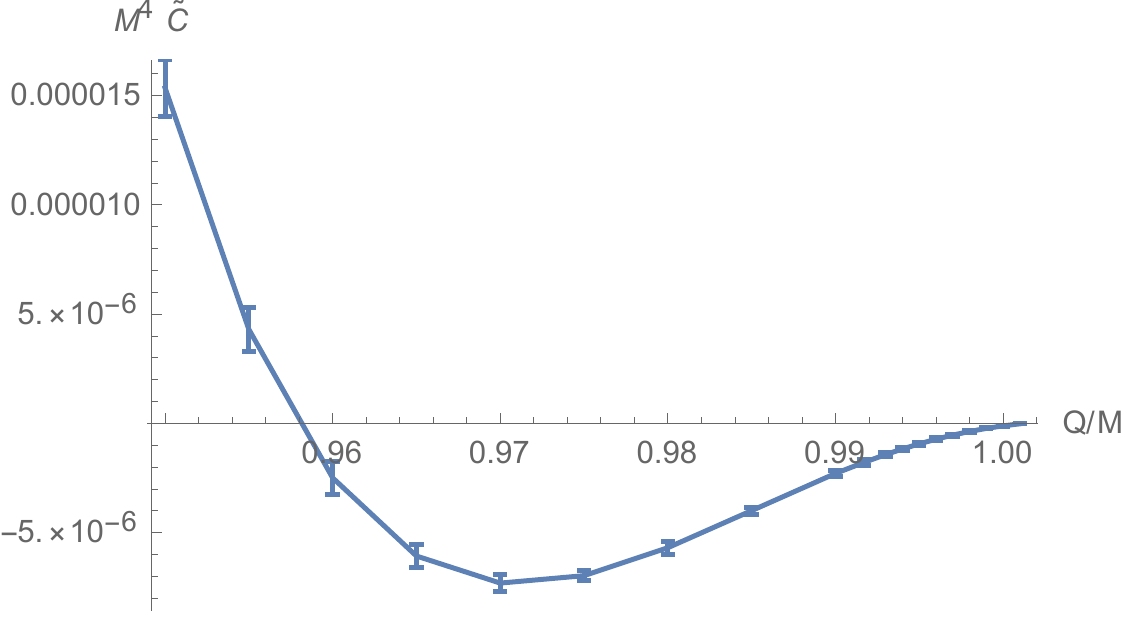}
\caption{\footnotesize$\langle T_{vv}\rangle \sim \tilde C$ at the Cauchy horizon in RNdS for a real massless scalar field. The cosmological constant is fixed to $\Lambda=0.02M^{-2}$. $\langle T_{vv}\rangle \sim \tilde C$ is plotted as a function of $Q/M$ in a broader range of the physical parameter regime.}
\label{fig:QvarWide}
\end{figure}

The values for the black-hole parameters are chosen such that they correspond to the least critical RNdS black hole leading to classical sCC violation with $\Lambda = 0.02 /M^{-2}$ for the blue, $\Lambda = 0.06 /M^{-2}$ for the orange and $\Lambda = 0.14 /M^{-2}$ for the green line in figure \ref{fig:CplotCard1} respectively \cite{Cardoso:2017}.

The most important property we observe is that $\tilde{C}$ is indeed non-zero in general, even though it becomes very small when we go to large scalar field masses in this extremal regime. Moreover, $\tilde{C}$ can be of either sign, and even passes through zero for a fixed set of spacetime parameters when changing the mass of the scalar field. This means that while the stress-energy tensor seems to diverge as $V^{-2}$ generically near the Cauchy horizon, it is not fixed whether this divergence will be towards $+\infty$ or $-\infty$. This, in turn, can decide whether nearby geodesics approaching the horizon will be accelerated towards or away from each other, and hence whether observers of finite size will be destroyed by stretching or squeezing, see \cite{Zilberman:2019,Hollands:2019} for a similar discussion.

The behavior of $\tilde{C}$ as a function of $Q/M$, the charge of the black hole relative to its mass, is shown in the  figures~\ref{fig:CplotCard2}-\ref{fig:CplotCard4}. For these plots, we keep the three values of the relative cosmological constant $\Lambda M^2$ considered in \cite{Cardoso:2017} fixed. They show that indeed $|\tilde{C}|$ decreases when approaching $Q=M$. Note that the larger $\Lambda$ for fixed $M$, the larger the difference between $r_+$ and $r_-$ when $Q$ reaches $M$. In this regime, corresponding to the classically sCC-violating regime found in \cite{Cardoso:2017}, it is not possible to see from figure~(\ref{fig:CplotCard2}-\ref{fig:CplotCard4}) that the sign of $\tilde{C}$ can also be changed by varying only the spacetime parameters while keeping the scalar field mass fixed. 

That changes if we consider a wider range of spacetime parameters. Figure~\ref{fig:QvarWide} shows $\tilde{C}$ for the massless scalar field, $\Lambda=0.02 M^{-2}$ and $Q/M$ between $0.95$ and $1.001$. We observe that $\tilde{C}$ changes its sign even for a fixed scalar field mass when considering also smaller values for the black hole charge. Similar results were obtained in \cite{Zilberman:2019} for a massless scalar field in the Reissner-Nordstr\"om spacetime by different methods. Comparing our results to figure~1 of \cite{Zilberman:2019}, we see that our results show very similar features, including the change of sign of $\tilde{C}$, which corresponds to $T_{vv}^{\rU}$ in there. 

%

\section{Conclusion}
\label{sec:concl}
We have applied a semi-analytical procedure to obtain the quantum stress-energy tensor for a real scalar quantum field on Reissner-Nordstr\"om-de Sitter spacetimes in an arbitrary state which is Hadamard in a neighborhood of the initial Cauchy surface. For this, we have used that \cite{Hollands:2019} $\langle T_{VV}\rangle_\Psi\sim \kappa_-^{-2} \tilde{C} V^{-2}$, and calculated the constant $\tilde{C}$, which is expressed as a series in angular momentum quantum number, for a variety of scalar field masses and spacetime parameters. The results indicate that for small enough values of the scalar field mass and cosmological constant this series converges very rapidly. Moreover, we found $\tilde{C}$ to be generally non-zero in a regime where classically sCC is violated \cite{Cardoso:2017}. This confirms that the expected stress-energy tensor of the quantum scalar field generically diverges at the Cauchy horizon in a very bad way, saving in particular the sCC conjecture e.g.\ in the Christodoulou formulation. We also observed that $\tilde{C}$ can be of either sign. This is still true if either the mass of the scalar field or the spacetime parameters are varied while the other is held fixed. In summary, strong cosmic censorship seems to be restored by quantum effects in general, and no observer can cross the Cauchy horizon. Whether the observer will be destroyed by stretching or squeezing depends on the sign of $\tilde{C}$ and thus on the parameters of the spacetime and of the the field theory. 

One way or another, it is physically very surprising -- but apparently true -- that quantum effects should be so large compared to the classical ones in this situation. 

\section*{Acknowledgements}
This work has been funded by the Deutsche Forschungsgemeinschaft (DFG) under the Grant No. 406116891 within the Research Training Group RTG 2522/1.

\bibliography{num_calc}

\begin{thebibliography}{28}%
\makeatletter
\providecommand \@ifxundefined [1]{%
 \@ifx{#1\undefined}
}%
\providecommand \@ifnum [1]{%
 \ifnum #1\expandafter \@firstoftwo
 \else \expandafter \@secondoftwo
 \fi
}%
\providecommand \@ifx [1]{%
 \ifx #1\expandafter \@firstoftwo
 \else \expandafter \@secondoftwo
 \fi
}%
\providecommand \natexlab [1]{#1}%
\providecommand \enquote  [1]{``#1''}%
\providecommand \bibnamefont  [1]{#1}%
\providecommand \bibfnamefont [1]{#1}%
\providecommand \citenamefont [1]{#1}%
\providecommand \href@noop [0]{\@secondoftwo}%
\providecommand \href [0]{\begingroup \@sanitize@url \@href}%
\providecommand \@href[1]{\@@startlink{#1}\@@href}%
\providecommand \@@href[1]{\endgroup#1\@@endlink}%
\providecommand \@sanitize@url [0]{\catcode `\\12\catcode `\$12\catcode
  `\&12\catcode `\#12\catcode `\^12\catcode `\_12\catcode `\%12\relax}%
\providecommand \@@startlink[1]{}%
\providecommand \@@endlink[0]{}%
\providecommand \url  [0]{\begingroup\@sanitize@url \@url }%
\providecommand \@url [1]{\endgroup\@href {#1}{\urlprefix }}%
\providecommand \urlprefix  [0]{URL }%
\providecommand \Eprint [0]{\href }%
\providecommand \doibase [0]{http://dx.doi.org/}%
\providecommand \selectlanguage [0]{\@gobble}%
\providecommand \bibinfo  [0]{\@secondoftwo}%
\providecommand \bibfield  [0]{\@secondoftwo}%
\providecommand \translation [1]{[#1]}%
\providecommand \BibitemOpen [0]{}%
\providecommand \bibitemStop [0]{}%
\providecommand \bibitemNoStop [0]{.\EOS\space}%
\providecommand \EOS [0]{\spacefactor3000\relax}%
\providecommand \BibitemShut  [1]{\csname bibitem#1\endcsname}%
\let\auto@bib@innerbib\@empty
\bibitem [{\citenamefont {Hollands}\ \emph {et~al.}(2020)\citenamefont
  {Hollands}, \citenamefont {Wald},\ and\ \citenamefont
  {Zahn}}]{Hollands:2019}%
  \BibitemOpen
  \bibfield  {author} {\bibinfo {author} {\bibfnamefont {S.}~\bibnamefont
  {Hollands}}, \bibinfo {author} {\bibfnamefont {R.~M.}\ \bibnamefont {Wald}},
  \ and\ \bibinfo {author} {\bibfnamefont {J.}~\bibnamefont {Zahn}},\ }\href
  {\doibase 10.1088/1361-6382/ab8052} {\bibfield  {journal} {\bibinfo
  {journal} {Class. Quant. Grav.}\ }\textbf {\bibinfo {volume} {37}},\ \bibinfo
  {pages} {115009} (\bibinfo {year} {2020})},\ \Eprint
  {http://arxiv.org/abs/1912.06047} {arXiv:1912.06047 [gr-qc]} \BibitemShut
  {NoStop}%
\bibitem [{\citenamefont {Christodoulou}(2009)}]{Christodoulou:2008}%
  \BibitemOpen
  \bibfield  {author} {\bibinfo {author} {\bibfnamefont {D.}~\bibnamefont
  {Christodoulou}},\ }\href@noop {} {\emph {\bibinfo {title} {{The Formation of
  Black Holes in General Relativity}}}}\ (\bibinfo  {publisher} {European
  Mathematical Society Publishing House},\ \bibinfo {address} {Zürich},\
  \bibinfo {year} {2009})\ \Eprint {http://arxiv.org/abs/0805.3880}
  {arXiv:0805.3880 [gr-qc]} \BibitemShut {NoStop}%
\bibitem [{\citenamefont {Dias}\ \emph {et~al.}(2018)\citenamefont {Dias},
  \citenamefont {Reall},\ and\ \citenamefont {Santos}}]{Dias:2018}%
  \BibitemOpen
  \bibfield  {author} {\bibinfo {author} {\bibfnamefont {O.~J.}\ \bibnamefont
  {Dias}}, \bibinfo {author} {\bibfnamefont {H.~S.}\ \bibnamefont {Reall}}, \
  and\ \bibinfo {author} {\bibfnamefont {J.~E.}\ \bibnamefont {Santos}},\
  }\href {\doibase 10.1007/JHEP10(2018)001} {\bibfield  {journal} {\bibinfo
  {journal} {JHEP}\ }\textbf {\bibinfo {volume} {10}},\ \bibinfo {pages} {001}
  (\bibinfo {year} {2018})},\ \Eprint {http://arxiv.org/abs/1808.02895}
  {arXiv:1808.02895 [gr-qc]} \BibitemShut {NoStop}%
\bibitem [{\citenamefont {Penrose}(1974)}]{Penrose:1974}%
  \BibitemOpen
  \bibfield  {author} {\bibinfo {author} {\bibfnamefont {R.}~\bibnamefont
  {Penrose}},\ }\enquote {\bibinfo {title} {Gravitational radiation and
  gravitational collapse},}\ \ (\bibinfo  {publisher} {Springer},\ \bibinfo
  {address} {Heidelberg},\ \bibinfo {year} {1974})\ Chap.\ \bibinfo {chapter}
  {Gravitational collapse}\BibitemShut {NoStop}%
\bibitem [{\citenamefont {Poisson}\ and\ \citenamefont
  {Israel}(1989)}]{Poisson:1989}%
  \BibitemOpen
  \bibfield  {author} {\bibinfo {author} {\bibfnamefont {E.}~\bibnamefont
  {Poisson}}\ and\ \bibinfo {author} {\bibfnamefont {W.}~\bibnamefont
  {Israel}},\ }\href {\doibase 10.1103/PhysRevLett.63.1663} {\bibfield
  {journal} {\bibinfo  {journal} {Phys. Rev. Lett.}\ }\textbf {\bibinfo
  {volume} {63}},\ \bibinfo {pages} {1663} (\bibinfo {year}
  {1989})}\BibitemShut {NoStop}%
\bibitem [{\citenamefont {Poisson}\ and\ \citenamefont
  {Israel}(1990)}]{Poisson:1990}%
  \BibitemOpen
  \bibfield  {author} {\bibinfo {author} {\bibfnamefont {E.}~\bibnamefont
  {Poisson}}\ and\ \bibinfo {author} {\bibfnamefont {W.}~\bibnamefont
  {Israel}},\ }\href {\doibase 10.1103/PhysRevD.41.1796} {\bibfield  {journal}
  {\bibinfo  {journal} {Phys.\ Rev.\ D}\ }\textbf {\bibinfo {volume} {41}},\
  \bibinfo {pages} {1796} (\bibinfo {year} {1990})}\BibitemShut {NoStop}%
\bibitem [{\citenamefont {Brady}\ \emph {et~al.}(1998)\citenamefont {Brady},
  \citenamefont {Moss},\ and\ \citenamefont {Myers}}]{Brady:1998}%
  \BibitemOpen
  \bibfield  {author} {\bibinfo {author} {\bibfnamefont {P.~R.}\ \bibnamefont
  {Brady}}, \bibinfo {author} {\bibfnamefont {I.~G.}\ \bibnamefont {Moss}}, \
  and\ \bibinfo {author} {\bibfnamefont {R.~C.}\ \bibnamefont {Myers}},\ }\href
  {\doibase 10.1103/PhysRevLett.80.3432} {\bibfield  {journal} {\bibinfo
  {journal} {Phys.\ Rev.\ Lett.}\ }\textbf {\bibinfo {volume} {80}},\ \bibinfo
  {pages} {3432} (\bibinfo {year} {1998})},\ \Eprint
  {http://arxiv.org/abs/gr-qc/9801032} {arXiv:gr-qc/9801032} \BibitemShut
  {NoStop}%
\bibitem [{\citenamefont {Mellor}\ and\ \citenamefont
  {Moss}(1990)}]{Mellor:1989}%
  \BibitemOpen
  \bibfield  {author} {\bibinfo {author} {\bibfnamefont {F.}~\bibnamefont
  {Mellor}}\ and\ \bibinfo {author} {\bibfnamefont {I.}~\bibnamefont {Moss}},\
  }\href {\doibase 10.1103/PhysRevD.41.403} {\bibfield  {journal} {\bibinfo
  {journal} {Phys.\ Rev.\ D}\ }\textbf {\bibinfo {volume} {41}},\ \bibinfo
  {pages} {403} (\bibinfo {year} {1990})}\BibitemShut {NoStop}%
\bibitem [{\citenamefont {Mellor}\ and\ \citenamefont
  {Moss}(1992)}]{Mellor:1992}%
  \BibitemOpen
  \bibfield  {author} {\bibinfo {author} {\bibfnamefont {F.}~\bibnamefont
  {Mellor}}\ and\ \bibinfo {author} {\bibfnamefont {I.}~\bibnamefont {Moss}},\
  }\href {\doibase 10.1088/0264-9381/9/4/001} {\bibfield  {journal} {\bibinfo
  {journal} {Classical and Quantum Gravity}\ }\textbf {\bibinfo {volume} {9}},\
  \bibinfo {pages} {L43} (\bibinfo {year} {1992})}\BibitemShut {NoStop}%
\bibitem [{\citenamefont {Dafermos}(2003)}]{Dafermos:2003a}%
  \BibitemOpen
  \bibfield  {author} {\bibinfo {author} {\bibfnamefont {M.}~\bibnamefont
  {Dafermos}},\ }\href@noop {} {\bibfield  {journal} {\bibinfo  {journal} {Ann.
  Math}\ }\textbf {\bibinfo {volume} {158}},\ \bibinfo {pages} {875} (\bibinfo
  {year} {2003})}\BibitemShut {NoStop}%
\bibitem [{\citenamefont {Dafermos}(2005)}]{Dafermos:2003b}%
  \BibitemOpen
  \bibfield  {author} {\bibinfo {author} {\bibfnamefont {M.}~\bibnamefont
  {Dafermos}},\ }\href@noop {} {\bibfield  {journal} {\bibinfo  {journal}
  {Commun. Pure Appl. Math.}\ }\textbf {\bibinfo {volume} {58}},\ \bibinfo
  {pages} {0445} (\bibinfo {year} {2005})},\ \Eprint
  {http://arxiv.org/abs/gr-qc/0307013} {arXiv:gr-qc/0307013} \BibitemShut
  {NoStop}%
\bibitem [{\citenamefont {Dafermos}(2014)}]{Dafermos:2012}%
  \BibitemOpen
  \bibfield  {author} {\bibinfo {author} {\bibfnamefont {M.}~\bibnamefont
  {Dafermos}},\ }\href {\doibase 10.1007/s00220-014-2063-4} {\bibfield
  {journal} {\bibinfo  {journal} {Commun. Math. Phys.}\ }\textbf {\bibinfo
  {volume} {332}},\ \bibinfo {pages} {729} (\bibinfo {year} {2014})},\ \Eprint
  {http://arxiv.org/abs/1201.1797} {arXiv:1201.1797 [gr-qc]} \BibitemShut
  {NoStop}%
\bibitem [{\citenamefont {Costa}\ \emph
  {et~al.}(2015{\natexlab{a}})\citenamefont {Costa}, \citenamefont {Girão},
  \citenamefont {Natário},\ and\ \citenamefont {Silva}}]{Costa:2014a}%
  \BibitemOpen
  \bibfield  {author} {\bibinfo {author} {\bibfnamefont {J.~L.}\ \bibnamefont
  {Costa}}, \bibinfo {author} {\bibfnamefont {P.~M.}\ \bibnamefont {Girão}},
  \bibinfo {author} {\bibfnamefont {J.}~\bibnamefont {Natário}}, \ and\
  \bibinfo {author} {\bibfnamefont {J.~D.}\ \bibnamefont {Silva}},\ }\href
  {\doibase 10.1088/0264-9381/32/1/015017} {\bibfield  {journal} {\bibinfo
  {journal} {Class. Quant. Grav.}\ }\textbf {\bibinfo {volume} {32}},\ \bibinfo
  {pages} {015017} (\bibinfo {year} {2015}{\natexlab{a}})},\ \Eprint
  {http://arxiv.org/abs/1406.7245} {arXiv:1406.7245 [gr-qc]} \BibitemShut
  {NoStop}%
\bibitem [{\citenamefont {Costa}\ \emph
  {et~al.}(2015{\natexlab{b}})\citenamefont {Costa}, \citenamefont {Girão},
  \citenamefont {Natário},\ and\ \citenamefont {Silva}}]{Costa:2014b}%
  \BibitemOpen
  \bibfield  {author} {\bibinfo {author} {\bibfnamefont {J.~L.}\ \bibnamefont
  {Costa}}, \bibinfo {author} {\bibfnamefont {P.~M.}\ \bibnamefont {Girão}},
  \bibinfo {author} {\bibfnamefont {J.}~\bibnamefont {Natário}}, \ and\
  \bibinfo {author} {\bibfnamefont {J.~D.}\ \bibnamefont {Silva}},\ }\href
  {\doibase 10.1007/s00220-015-2433-6} {\bibfield  {journal} {\bibinfo
  {journal} {Commun. Math. Phys.}\ }\textbf {\bibinfo {volume} {339}},\
  \bibinfo {pages} {903} (\bibinfo {year} {2015}{\natexlab{b}})},\ \Eprint
  {http://arxiv.org/abs/1406.7253} {arXiv:1406.7253 [gr-qc]} \BibitemShut
  {NoStop}%
\bibitem [{\citenamefont {Costa}\ \emph {et~al.}(2017)\citenamefont {Costa},
  \citenamefont {Girão}, \citenamefont {Natário},\ and\ \citenamefont
  {Silva}}]{Costa:2014c}%
  \BibitemOpen
  \bibfield  {author} {\bibinfo {author} {\bibfnamefont {J.~L.}\ \bibnamefont
  {Costa}}, \bibinfo {author} {\bibfnamefont {P.~M.}\ \bibnamefont {Girão}},
  \bibinfo {author} {\bibfnamefont {J.}~\bibnamefont {Natário}}, \ and\
  \bibinfo {author} {\bibfnamefont {J.~D.}\ \bibnamefont {Silva}},\ }\href
  {\doibase 10.1007/s40818-017-0028-6} {\bibfield  {journal} {\bibinfo
  {journal} {Annals of PDE}\ }\textbf {\bibinfo {volume} {3}} (\bibinfo {year}
  {2017}),\ 10.1007/s40818-017-0028-6},\ \Eprint
  {http://arxiv.org/abs/1406.7261} {arXiv:1406.7261 [gr-qc]} \BibitemShut
  {NoStop}%
\bibitem [{\citenamefont {Franzen}(2016)}]{Franzen:2014}%
  \BibitemOpen
  \bibfield  {author} {\bibinfo {author} {\bibfnamefont {A.~T.}\ \bibnamefont
  {Franzen}},\ }\href {\doibase 10.1007/s00220-015-2440-7} {\bibfield
  {journal} {\bibinfo  {journal} {Commun. Math. Phys.}\ }\textbf {\bibinfo
  {volume} {343}},\ \bibinfo {pages} {601} (\bibinfo {year} {2016})},\ \Eprint
  {http://arxiv.org/abs/1407.7093} {arXiv:1407.7093 [gr-qc]} \BibitemShut
  {NoStop}%
\bibitem [{\citenamefont {Luk}\ and\ \citenamefont {Oh}(2017)}]{Luk:2015}%
  \BibitemOpen
  \bibfield  {author} {\bibinfo {author} {\bibfnamefont {J.}~\bibnamefont
  {Luk}}\ and\ \bibinfo {author} {\bibfnamefont {S.-J.}\ \bibnamefont {Oh}},\
  }\href {\doibase 10.1215/00127094-3715189} {\bibfield  {journal} {\bibinfo
  {journal} {Duke Math.\ J.}\ }\textbf {\bibinfo {volume} {166}},\ \bibinfo
  {pages} {437} (\bibinfo {year} {2017})},\ \Eprint
  {http://arxiv.org/abs/1501.04598} {arXiv:1501.04598 [gr-qc]} \BibitemShut
  {NoStop}%
\bibitem [{\citenamefont {Hintz}\ and\ \citenamefont
  {Vasy}(2017)}]{Hintz:2015}%
  \BibitemOpen
  \bibfield  {author} {\bibinfo {author} {\bibfnamefont {P.}~\bibnamefont
  {Hintz}}\ and\ \bibinfo {author} {\bibfnamefont {A.}~\bibnamefont {Vasy}},\
  }\href {\doibase 10.1063/1.4996575} {\bibfield  {journal} {\bibinfo
  {journal} {J.\ Math.\ Phys.}\ }\textbf {\bibinfo {volume} {58}},\ \bibinfo
  {pages} {081509} (\bibinfo {year} {2017})},\ \Eprint
  {http://arxiv.org/abs/1512.08004} {arXiv:1512.08004 [math.AP]} \BibitemShut
  {NoStop}%
\bibitem [{\citenamefont {Costa}\ and\ \citenamefont
  {Franzen}(2017)}]{Costa:2016}%
  \BibitemOpen
  \bibfield  {author} {\bibinfo {author} {\bibfnamefont {J.~L.}\ \bibnamefont
  {Costa}}\ and\ \bibinfo {author} {\bibfnamefont {A.~T.}\ \bibnamefont
  {Franzen}},\ }\href {\doibase 10.1007/s00023-017-0592-z} {\bibfield
  {journal} {\bibinfo  {journal} {Annales Henri Poincare}\ }\textbf {\bibinfo
  {volume} {18}},\ \bibinfo {pages} {3371} (\bibinfo {year} {2017})},\ \Eprint
  {http://arxiv.org/abs/1607.01018} {arXiv:1607.01018 [gr-qc]} \BibitemShut
  {NoStop}%
\bibitem [{\citenamefont {Cardoso}\ \emph
  {et~al.}(2018{\natexlab{a}})\citenamefont {Cardoso}, \citenamefont {Costa},
  \citenamefont {Destounis}, \citenamefont {Hintz},\ and\ \citenamefont
  {Jansen}}]{Cardoso:2017}%
  \BibitemOpen
  \bibfield  {author} {\bibinfo {author} {\bibfnamefont {V.}~\bibnamefont
  {Cardoso}}, \bibinfo {author} {\bibfnamefont {J.~a.~L.}\ \bibnamefont
  {Costa}}, \bibinfo {author} {\bibfnamefont {K.}~\bibnamefont {Destounis}},
  \bibinfo {author} {\bibfnamefont {P.}~\bibnamefont {Hintz}}, \ and\ \bibinfo
  {author} {\bibfnamefont {A.}~\bibnamefont {Jansen}},\ }\href {\doibase
  10.1103/PhysRevLett.120.031103} {\bibfield  {journal} {\bibinfo  {journal}
  {Phys.\ Rev.\ Lett.}\ }\textbf {\bibinfo {volume} {120}},\ \bibinfo {pages}
  {031103} (\bibinfo {year} {2018}{\natexlab{a}})},\ \Eprint
  {http://arxiv.org/abs/1711.10502} {arXiv:1711.10502 [gr-qc]} \BibitemShut
  {NoStop}%
\bibitem [{\citenamefont {Hod}(2019)}]{Hod:2018}%
  \BibitemOpen
  \bibfield  {author} {\bibinfo {author} {\bibfnamefont {S.}~\bibnamefont
  {Hod}},\ }\href {\doibase 10.1016/j.nuclphysb.2019.03.003} {\bibfield
  {journal} {\bibinfo  {journal} {Nucl. Phys. B}\ }\textbf {\bibinfo {volume}
  {941}},\ \bibinfo {pages} {636} (\bibinfo {year} {2019})},\ \Eprint
  {http://arxiv.org/abs/1801.07261} {arXiv:1801.07261 [gr-qc]} \BibitemShut
  {NoStop}%
\bibitem [{\citenamefont {Dias}\ \emph {et~al.}(2019)\citenamefont {Dias},
  \citenamefont {Reall},\ and\ \citenamefont {Santos}}]{Dias:2018a}%
  \BibitemOpen
  \bibfield  {author} {\bibinfo {author} {\bibfnamefont {O.~J.}\ \bibnamefont
  {Dias}}, \bibinfo {author} {\bibfnamefont {H.~S.}\ \bibnamefont {Reall}}, \
  and\ \bibinfo {author} {\bibfnamefont {J.~E.}\ \bibnamefont {Santos}},\
  }\href {\doibase 10.1088/1361-6382/aafcf2} {\bibfield  {journal} {\bibinfo
  {journal} {Class. Quant. Grav.}\ }\textbf {\bibinfo {volume} {36}},\ \bibinfo
  {pages} {045005} (\bibinfo {year} {2019})},\ \Eprint
  {http://arxiv.org/abs/1808.04832} {arXiv:1808.04832 [gr-qc]} \BibitemShut
  {NoStop}%
\bibitem [{\citenamefont {Cardoso}\ \emph
  {et~al.}(2018{\natexlab{b}})\citenamefont {Cardoso}, \citenamefont {Costa},
  \citenamefont {Destounis}, \citenamefont {Hintz},\ and\ \citenamefont
  {Jansen}}]{Cardoso:2018}%
  \BibitemOpen
  \bibfield  {author} {\bibinfo {author} {\bibfnamefont {V.}~\bibnamefont
  {Cardoso}}, \bibinfo {author} {\bibfnamefont {J.~L.}\ \bibnamefont {Costa}},
  \bibinfo {author} {\bibfnamefont {K.}~\bibnamefont {Destounis}}, \bibinfo
  {author} {\bibfnamefont {P.}~\bibnamefont {Hintz}}, \ and\ \bibinfo {author}
  {\bibfnamefont {A.}~\bibnamefont {Jansen}},\ }\href {\doibase
  10.1103/PhysRevD.98.104007} {\bibfield  {journal} {\bibinfo  {journal}
  {Phys.\ Rev.\ D}\ }\textbf {\bibinfo {volume} {98}},\ \bibinfo {pages}
  {104007} (\bibinfo {year} {2018}{\natexlab{b}})},\ \Eprint
  {http://arxiv.org/abs/1808.03631} {arXiv:1808.03631 [gr-qc]} \BibitemShut
  {NoStop}%
\bibitem [{\citenamefont {Mo}\ \emph {et~al.}(2018)\citenamefont {Mo},
  \citenamefont {Tian}, \citenamefont {Wang}, \citenamefont {Zhang},\ and\
  \citenamefont {Zhong}}]{Mo:2018}%
  \BibitemOpen
  \bibfield  {author} {\bibinfo {author} {\bibfnamefont {Y.}~\bibnamefont
  {Mo}}, \bibinfo {author} {\bibfnamefont {Y.}~\bibnamefont {Tian}}, \bibinfo
  {author} {\bibfnamefont {B.}~\bibnamefont {Wang}}, \bibinfo {author}
  {\bibfnamefont {H.}~\bibnamefont {Zhang}}, \ and\ \bibinfo {author}
  {\bibfnamefont {Z.}~\bibnamefont {Zhong}},\ }\href {\doibase
  10.1103/PhysRevD.98.124025} {\bibfield  {journal} {\bibinfo  {journal} {Phys.
  Rev. D}\ }\textbf {\bibinfo {volume} {98}},\ \bibinfo {pages} {124025}
  (\bibinfo {year} {2018})},\ \Eprint {http://arxiv.org/abs/1808.03635}
  {arXiv:1808.03635 [gr-qc]} \BibitemShut {NoStop}%
\bibitem [{\citenamefont {Gim}\ and\ \citenamefont {Gwak}(2019)}]{Gim:2019}%
  \BibitemOpen
  \bibfield  {author} {\bibinfo {author} {\bibfnamefont {Y.}~\bibnamefont
  {Gim}}\ and\ \bibinfo {author} {\bibfnamefont {B.}~\bibnamefont {Gwak}},\
  }\href {\doibase 10.1103/PhysRevD.100.124001} {\bibfield  {journal} {\bibinfo
   {journal} {Phys. Rev. D}\ }\textbf {\bibinfo {volume} {100}},\ \bibinfo
  {pages} {124001} (\bibinfo {year} {2019})},\ \Eprint
  {http://arxiv.org/abs/1901.11214} {arXiv:1901.11214 [gr-qc]} \BibitemShut
  {NoStop}%
\bibitem [{\citenamefont {Zilberman}\ \emph {et~al.}(2020)\citenamefont
  {Zilberman}, \citenamefont {Levi},\ and\ \citenamefont
  {Ori}}]{Zilberman:2019}%
  \BibitemOpen
  \bibfield  {author} {\bibinfo {author} {\bibfnamefont {N.}~\bibnamefont
  {Zilberman}}, \bibinfo {author} {\bibfnamefont {A.}~\bibnamefont {Levi}}, \
  and\ \bibinfo {author} {\bibfnamefont {A.}~\bibnamefont {Ori}},\ }\href
  {\doibase 10.1103/PhysRevLett.124.171302} {\bibfield  {journal} {\bibinfo
  {journal} {Phys. Rev. Lett.}\ }\textbf {\bibinfo {volume} {124}},\ \bibinfo
  {pages} {171302} (\bibinfo {year} {2020})},\ \Eprint
  {http://arxiv.org/abs/1906.11303} {arXiv:1906.11303 [gr-qc]} \BibitemShut
  {NoStop}%
\bibitem [{\citenamefont {Suzuki}\ \emph {et~al.}(1999)\citenamefont {Suzuki},
  \citenamefont {Takasugi},\ and\ \citenamefont {Umetsu}}]{Suzuki:1999}%
  \BibitemOpen
  \bibfield  {author} {\bibinfo {author} {\bibfnamefont {H.}~\bibnamefont
  {Suzuki}}, \bibinfo {author} {\bibfnamefont {E.}~\bibnamefont {Takasugi}}, \
  and\ \bibinfo {author} {\bibfnamefont {H.}~\bibnamefont {Umetsu}},\ }\href
  {\doibase 10.1143/PTP.102.253} {\bibfield  {journal} {\bibinfo  {journal}
  {Prog.\ Theor.\ Phys.}\ }\textbf {\bibinfo {volume} {102}},\ \bibinfo {pages}
  {253} (\bibinfo {year} {1999})},\ \Eprint
  {http://arxiv.org/abs/gr-qc/9905040} {arXiv:gr-qc/9905040} \BibitemShut
  {NoStop}%
\bibitem [{\citenamefont {Ronveaux}(1995)}]{Ronveaux:1995}%
  \BibitemOpen
  \bibinfo {editor} {\bibfnamefont {A.}~\bibnamefont {Ronveaux}},\ ed.,\
  \href@noop {} {\emph {\bibinfo {title} {Heun's differential equation}}}\
  (\bibinfo  {publisher} {Oxford University Press},\ \bibinfo {address}
  {Oxford},\ \bibinfo {year} {1995})\BibitemShut {NoStop}%
\end{thebibliography}%

\end{document}